\documentclass[sigconf]{acmart}

\usepackage{booktabs} 

\setcopyright{none}

\acmConference[arXiv]{arXiv}{Jan 2019}{}
\acmYear{2019}
\copyrightyear{2019}

\acmArticle{0}
\acmPrice{00.00}

\begin{document}
\title{Integrating and querying similar tables from PDF documents using deep learning}

\author{Rahul Anand}
\affiliation{%
  \institution{University of New South Wales}
  \city{Sydney}
  \country{Australia}
}
\email{rahulanand.inbox@gmail.com}

\author{Hye-young Paik}
\authornote{This work was done while the author was on sabbatical at Data61, CSIRO}
\orcid{0000-0003-4425-7388}
\affiliation{%
  \institution{University of New South Wales}
  \city{Sydney}
  \country{Australia}
}
\email{h.paik@unsw.edu.au}

\author{Chen Wang}
\affiliation{%
  \institution{Data61, CSIRO}
  \city{Sydney}
  \country{Australia}
}
\email{chen.wang@data61.csiro.au}





\renewcommand{\shortauthors}{R. Anand et al.}

\begin{abstract}
Large amount of public data produced by enterprises are in semi-structured PDF form. Tabular data extraction from reports and other published data in PDF format is of interest for various data consolidation purposes such as analysing and aggregating financial reports of a company.  Queries into the structured  tabular data in PDF format are normally processed in an unstructured manner through means like text-match. This is mainly due to that the binary format of PDF documents is optimized for layout and rendering and do not have great support for automated parsing of data. Moreover, even the same table type in PDF files varies in schema, row or column headers, which makes it difficult for a query plan to cover all relevant tables. This paper proposes a deep learning based method to enable SQL-like query and analysis of financial tables from annual reports in PDF format. This is achieved through table type classification and nearest row search. We demonstrate that using word embedding trained on Google news for header match clearly outperforms the text-match based approach in traditional database. We also introduce a practical system that uses this technology to query and analyse finance tables in PDF documents from various sources.    
\end{abstract}

%
%



\maketitle

\section{Introduction}

Tables in documents are an under-utilised source of data due to the ineffectiveness of current technologies in extracting the content in a useful manner \cite{PerezArriagaEA16,Texus}. However, documents often contain data that is not available elsewhere and due to its portable nature, more and more documents are used as a means of information exchange, which makes it necessary to develop an easy-to-use solution to access, integrate and query such data. 

PDF is the most common format of documents, especially for dissemination of data on the Web. Although the text-based format is machine-readable, there are no markups indicating the structure or layout of a document. It is formed as objects which encapsulate positional and styling information rather than the semantics of the content. Also, tables containing similar data (e.g., financial statements) may have drastically different layouts in different documents depending on their authors. These present a challenge for extracting tables embedded within a PDF document with high enough accuracy that can be used for data integration and analysis \cite{PerezArriagaEA16,khusro2014methods}. 
Queries into the structured tabular data present in PDF documents are normally processed in an unstructured manner through means like text-match.

In this paper, we look into financial documents an example domain. Tabular data extraction from reports and other published data in PDF format is of interest for various data consolidation purposes such as analysing and aggregating financial reports of a company.

We describe a methodology and prototype of an application that provides quick browsing and querying access to similar table data across many PDF files. The application also allows further analysis of the table content by querying similar rows and columns across the same class of ``similar'' tables. Through this application, we aim to demonstrate an approach for rapid integration of similar tables extracted from different PDF files.


There are projects and tools aimed at conversion of PDF documents into more structured markup formats while preserving the semantics of structures like tables \cite{Texus,Tegra}. Many of these tools focus on processing limited types of tables (e.g., completely ruled tables) from limited domains (e.g., scientific papers). Many also require human intervention or pre-determined target schema  \cite{TexusAuto}.  What we propose is not aimed at producing a highly precise data extraction technique with a strict schema, rather, we describe a complementary idea whereby a deep-learning based processing pipeline provides an automated experience for users at detecting tables, finding similar tables in the document set, performing semi-structured query analysis on the found tables. 

This is of course not a precise process, but nevertheless produces usable results for ``fuzzy'' search and analysis of the data which otherwise would not have been accessible. This project aims at laying the groundwork for a pipeline for obtaining useful insights from the PDF documents while avoiding manual effort to parse and classify the tables.

\section{Related Work}

We discuss related work from three view points. First, we provide an overview of table extraction and processing work, then we will discuss the recent body of work on utilising machine learning algorithms. In our own approach, we map PDF files to HTML so that the embedded tables are marked with HTML tags. In this regard, we also consider the work done in processing ``Web'' tables and its applications.


\paragraph{\underline{Table Extraction}} Extracting tables from documents require locating (recognising) tables from given document elements, then identifying individual cells that make up a table. Many different approaches have been proposed in the past, both from academia and commercial tool builders. However, the problem of extracting tables is still open and actively being investigated~\cite{khusro2014methods,TabNet17} by the research communities. 
Different approaches have been utilised to solve the problem in specific document formats and various application contexts. For example, techniques involved in extracting tables with HTML markups in Web pages~\cite{Pinto:2003,embley2016converting,Adelfio:2013} are different to recognising tables from scanned documents (i.e., images)~\cite{TranTNLYK16,GilaniQMS17,HeCPKG17}. Because PDF documents have no specific table  markups, text-based PDF documents~\cite{pdf2table05,oro2009pdf,Texus} require parsing techniques that are customised for processing the raw PDF format.  Also, the solutions for table extraction tasks tend to be devised in a particular application context. For example, a main body of work originated from the Web Table Corpora \cite{LehmbergRMB16} focus on knowledge base construction, matching table instances to concepts in DBPedia or Freebase \cite{RitzeB17}. Others are driven by the need for extracting medical/scientific data from formal scientific papers~\cite{constantin2013pdfx}, or financial documents\cite{LiSF16}.  Because the problem and application requirements are diverse, most solutions with regard to table extraction are often limited in scope. In particular, the text-based PDF processing work suffers from a lack of standarised schema for PDF parsers, leading to even more limited applicability. Our approach in this paper does not rely on proprietary or custom PDF parsing tools, but use the common Adobe PDF tool\footnote{Adobe Document Cloud, \url{https://acrobat.adobe.com/au/en/}} to convert PDF to HTML. We believe there is an opportunity to improve the utility of the text-based PDF table extraction efforts by moving towards a standard representation of the text schema such as HTML.


\paragraph{\underline{Machine Learning Approaches}}

A common approach to solving table extraction problem is to use heuristically designed process, based on observed patterns in the tables. Typically, those approaches \cite{pdf2table05} focused on the detecting line arts as the boundaries of tables, analysing coordinates and alignments of the text elements, or searching for already-known column headings, etc. 


Another approach is to use machine learning algorithms. Although this approach has been known in the table processing research \cite{khusro2014methods}, the recent advances in deep learning algorithms and readily available tools made the application of the technique more attractive. 



Some of the previous work on table detection with deep learning techniques involve detection of table blocks and their extraction from document images using convolutional neural networks \cite{Azka,Dafang}. These studies primarily work on extraction of tables from whole documents. 
In \cite{HaoGYT16}, Hao et al. proposed a convolutional neural networks based method for detecting tables from in PDF documents. They first identify the table-like areas using some initial rules and the convolutional network utilises the visual features of the chosen areas.  A similar deep learning approach is also presented in \cite{GilaniQMS17}.

A system called TAO in \cite{PerezArriagaEA16} uses the k-nearest neighbour method and pre-observed layout heuristics to detect tables and extract table cells from PDF. In \cite{TranTNLYK16}, the authors presented an approach for identifying table regions, based on a shape (called Random Rotation Bounding Box). Using the shape, they detected the table regions, such as identification of text, non-text elements or ruled-line tables  in document images. 

A study on web tables uses a hybrid deep neural network architecture to classify different `shapes/layout' of HTML tables is presented in
\cite{TabNet17}. 

Some of these work report high accuracy in recall and precision. In all machine learning approaches, to achieve the high level accuracy, it is combined with some basic heuristics and pre-determined rules. 
In our approach, we do minimal pre-processing of the raw data and do not inject any rule-based knowledge into the processing pipeline. In our application, we demonstrate that without customised pre-processing on input data, we can classify tables based on the content, and identify similar rows and matching columns within a class.

\paragraph{\underline{Web (HTML) tables}} The HTML markups for tables have long been used for page layout purposes and only a subset of the ``tables'' actually contain tabular data \cite{CafarellaHLMYWW18}. While some of the early work focus on separating table data from layout instructions \cite{Pinto:2003,Chen:2000,CafarellaHWWZ08}, projects such as WDC \cite{LehmbergRMB16} provide a large scale Web table data classified from general purpose Web crawl data. Most of the Web table processing work fall into the following two categories: table search and knowledge base construction. A keyword-based technique which ranks extracted tables based on table content \cite{BalakrishnanHHL15} is the most common application. Recently Google Search incorporated table search into its search results \footnote{https://research.google.com/tables}. Annotating entities or concepts in tables using a generic purpose knowledge \cite{Limaye:2010} or extending a knowledge base using the data contained in Web tables \cite{DBpedia2013} is an important application of Web tables. Matching tables could be applied to complete or extend a table itself (e.g., a country table with population and capital city could have extra columns such as area total added) \cite{LEHMBERG2015159}. Many of the these techniques fundamentally leverage the relationships amongst rows and columns often with a machine learning approach.




\section{Preliminaries}

In this section, we first describe the data domain used for building the application. Then, the core machine learning concepts utilised in the deep learning pipeline are summarised briefly.

\subsection{Dataset}

\subsubsection{Annual Report}
An annual report is an aggregated document detailing the actions of a company in the previous year. The primary purpose of this report is to provide transparency of information to shareholders and other stakeholders regarding the company's financial performance and related decisions. 

In this study we are interested in annual reports from Australian companies. Corporate entities are required to prepare financial reports subject to legislative requirements applicable throughout all Australian states and territories. The regulatory requirements of annual reports are governed by the Australian Securities Exchange (ASX), the Australian Taxation Office (ATO) and the Australian Securities and Investments Commission (ASIC) 
\cite{CpaAuAnnualReports}. The standards primarily enforces four different types of financial statements, as described in below sections. These statements are mainly aimed at assisting the shareholders in planning their allotment of funds. The standard body which governs the preparation of financial statements is the Australian
Accounting Standards Board (AASB).

\begin{itemize}
    
\item{Statement of Financial Position (Balance Sheet)}
A statement of financial position is a brief summary of all company controlled resources, and all dues owed by the organisation at a given point in time. This includes the assets, liabilities, and equity related information. As it is recorded at a single point, it can be considered a snapshot of a company's financial capacity 

\item{Profit or Loss Statement (Income Statement)}
The profit or loss statement is a report recording events over a period showing how the revenues and expenses are carried out. This report serves as a reference for company administration and financial supporters to understand whether the company is running at a profit or at a loss.

\item{Statement of Changes in Equity}
The statement of changes in equity documents the changes in all of the organisation's equity throughout a financial period.

\item{Cash Flow Statement}
The cash flow statement reports flow of cash in and out of the company, spanning over operating, investing and financing activities.
\end{itemize}

Many parts of the annual reports are of interests to general public, as well as to the academic researchers who study financial regulations or trend analysis. Although these reports are publicly available, the extracted and curated data from the reports in a ready-to-use form tends to be behind a pay-wall and not easily accessible.

\subsection{Machine Learning Concepts}

\subsubsection{Word2vec}
Word2vec is a neural network model used to create word embeddings. The model is trained on a set of words first converted to their encoded representations. 
There are two representations that are primarily used with word2vec \cite{Word2vecMikolov}. \textit{Continuous bag of words (CBOW)} considers nearby words of a given word as a single set, so that the word order does not influence prediction. This representation enables prediction of a word from surrounding bag of words. In \textit{Skip-gram} model, the distance of nearby words influence the prediction. Skip-gram enables the prediction of nearby context words from a given word.

On training word2vec on the input vocabulary of words, the output is a mapping of the words in a multidimensional space. The peculiarity of the vector representing each word is that the data point is closer to similar words in the multidimensional space. The similarity measure is the contextual usage of words. 

\subsubsection{LSTM}
A recurrent neural network (RNN) is a neural network designed to process sequential data such as text content \cite{Goodfellow-et-al-2016}. 
Long short-term memory (LSTM) is a variant of RNN unit. These units can process and remember long-term dependencies among the input sequence of tokens \cite{ColahLSTM}. An LSTM cell is composed of a input, output, and forget gates. With the use of gates to control the incoming stream, the cell can remember information across time in its hidden state.

\subsubsection{Softmax}
The softmax function normalizes an N-dimensional vector of real numbers into a vector of real numbers in the range (0,1). This new vector can be seen as the probability distribution of N values, and thus the values add up to 1.

\subsubsection{K-Nearest Neighbour algorithm}
KNN is a supervised algorithm which uses a feature vector associated with a data point to compute geometric distances between samples, thereby positioning them in a multidimensional space. For a given test sample, K closest training examples are computed by the algorithm. In a classification problem, the majority class is assigned to the test sample.

\subsubsection{Nearest Neighbour clustering}
The idea of distance between sample points can be utilised to cluster a dataset where the number and type of clusters are unknown. This is achieved by considering the closest neighbours within a radius of the test sample with some cutoff. Partitioning data structures such as KD-tree or ball tree are used for fast clustering.

\subsubsection{t-SNE algorithm}
T-distributed Stochastic Neighbour Embedding is a dimensionality reduction algorithm used in visualisation of high dimensional spaces. Using t-SNE, a higher dimensional point is mapped to a point in 2 or 3 dimensional space, preserving the similarity of the sample points in their distance from each other in the visualisation space.

\section{Methodology}

In this section, we describe the deep learning methods designed to process tables. We will present an overview of the end-to-end processing pipeline. Then, we explain the classification process to identify similar tables. From there, we introduce our ``similarity'' query framework which returns similar rows and associated columns across different documents for a given table row as a query.

\subsection{Overview}

\begin{figure}
\includegraphics[width=\linewidth]{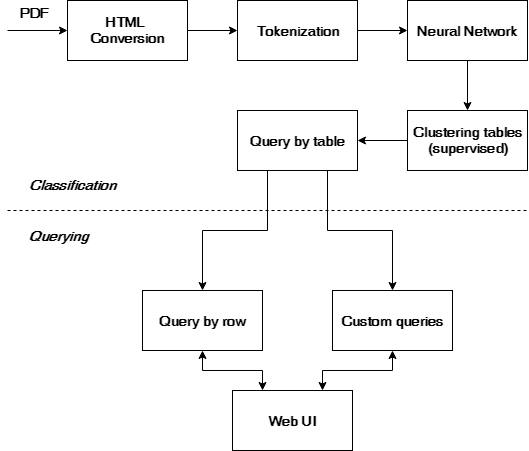}
\caption{End to end pipeline for ad-hoc PDF table analysis using deep learning}
\label{fig:pipeline_revised}
\end{figure}

Using publicly available annual reports as an example data source, we designed a pipeline which can assist in the automated extraction and identification of tables. The results are then made available through an interactive display for quick analysis. 

The current implementation can identify the following four types/classes of financial tables from a typical annual report: {\it Profit or Loss}, {\it Financial Position}, {\it Changes in Equity}, and {\it Cash flows}. 

Figure~\ref{fig:pipeline_revised} on page~\pageref{fig:pipeline_revised} shows the design of the end-to-end pipeline.   The first part of the processing line is to identify and classify tables into the four aforementioned table types. We  train a model on extracted tables containing the four types using a deep learning architecture. 

The second part of the processing line is to handle similarity queries over the extracted tables. There are two types of queries: querying by table, and querying by row/column. A query by table allows users to select a table and have similar tables returned. A query by rows/columns allow users to select a row in a table and have similar rows from other tables returned. This can be further refined using range operators. 
The prediction probability obtained from the classification phase is used as a distance vector for clustering  tables.

The resultant system demonstrates an architecture for automatically classifying and clustering any new tables of the given types from text-based PDF files. Further it enables querying a specific table and retrieving related tables of the same class and/or company. A machine learning based analysis of table rows from two candidate tables also enables quick comparison of the content. 

\subsection{Table Classification}
This section describes the first part of the processing pipeline which is for classification.

\begin{figure}[h]
\includegraphics[width=\linewidth]{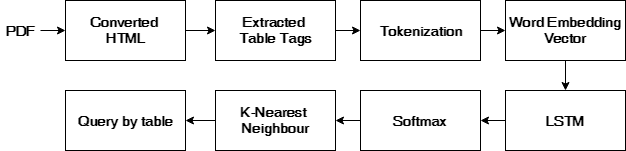}
\caption{Deep learning architecture for table classification}
\label{fig:classification_pipeline}
\end{figure}

\subsubsection{Data collection and pre-processing}
We have downloaded around 100 annual reports from various Australian companies published over multiple financial years from the Australian Stock Exchange Web site\footnote{ASX announcements, \url{https://www.asx.com.au/asx/statistics/announcements.do}}. 

A PDF file is a binary format devoid of any defined markup for elements like tables. The file is defined in terms of objects and their positioning which relates to the visual rendering. Hence working with a PDF file directly will require parsing of table structures based on heuristics. 

Instead of relying on a PDF parser that produces non-standard outputs, an intermediate step that is beneficial here is to convert the PDF file into HTML format using an industry standard conversion tool and extract table tags from the web format. The conversion, while noisy, is reliable for further processing. 

The extracted tables can then be annotated with appropriate class type and the name of the company that produced the table. We have manually extracted and labelled 
146 unique tables spanning across 5 companies. This formed the training and test data for the deep learning system for classification.

\subsubsection{Tokenization}

\begin{figure}
\includegraphics[width=\linewidth]{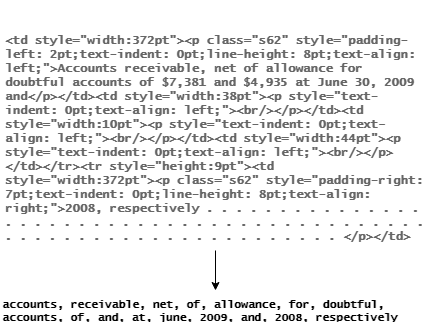}
\caption{Tokenization of a row from converted HTML table}
\label{fig:tokenization}
\end{figure}

We are primarily interested in being able to classify a given table as a standard financial table type, and also as belonging to a particular company (as the same table type can differ in terminology and structure across companies). The numerical values in rows except for the years are irrelevant here as they are unique values and do not distinguish one type of financial table in any way. We also cleanup the HTML tags surrounding the table cells, as the columnar structure of a given row is not relevant for our classification (See Figure \ref{fig:tokenization} on page \pageref{fig:tokenization}). Thus a table is represented as a stream of word tokens mostly comprising of row descriptions and column headers. 

\subsubsection{Classification}

\begin{figure}
\includegraphics[width=\linewidth]{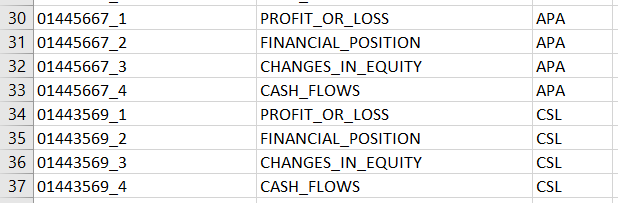}
\caption{Snippet of training data with labels. The first column refers to the file names pointing to extracted HTML tables. Second and third columns are standard table type and company name respectively. These are combined as required and mapped to integers for labelling the training data.}
\label{fig:classification_data}
\end{figure}

The tokenized tables along with their annotated labels are used as input for a neural network architecture for classification. Classes are formed by assigning a running ID to a combination of a company name and table type (See Figure \ref{fig:classification_data} on page \pageref{fig:classification_data}). 

The same architecture can be {\it reused} to train only on table types by setting the company name to a dummy entry for all samples. We proceed with the choice of training specifically to a company name and table type.

Each table is treated as a single sentence of tokens and is converted into a numerical vector using a word embedding. 

We tried different choices of word embeddings at this step. One method is to train a custom word2vec embedding on the vocabulary of tokens extracted from the entire set of tables in our dataset. The skip-gram model trained this way contained 870 unique words from the cleaned up tokens, represented as 100-dimensional vectors. Another method is using a pre-trained model from Google, {\it the Google News} word2vec model. This model contains 3 million words represented as 300-dimensional vectors, and is trained on 100 billion words from a Google News dataset\footnote{Google News dataset, \url{https://code.google.com/archive/p/word2vec}}. 

The tables converted to vectors with the choice of word embedding is then trained with LSTM cells to learn patterns from the tokenized tables. The choice of LSTM cells for training is desirable to learn the patterns in the vocabularies used in the table cell descriptors, and understand the positional structure and order of the words. 

Individual table entries are translated into vectors of size 40. This input vector is fed through an RNN formed by a stack of 4 LSTM cells, and the loss is computed after applying softmax to the resultant classification vector. We use Adam optimizer to minimize the loss. Furthermore, the normalized vector of class prediction probabilities obtained after softmax is recorded separately for each table (See Figure \ref{fig:classification_pipeline} on page \pageref{fig:classification_pipeline}). This data is utilised in the clustering step for queryig similar tables.

\subsection{Querying} 
We support two types of ``similarity'' queries: a query by table, a query by row. This section describes the query processing steps.
\begin{figure}
\includegraphics[width=\linewidth]{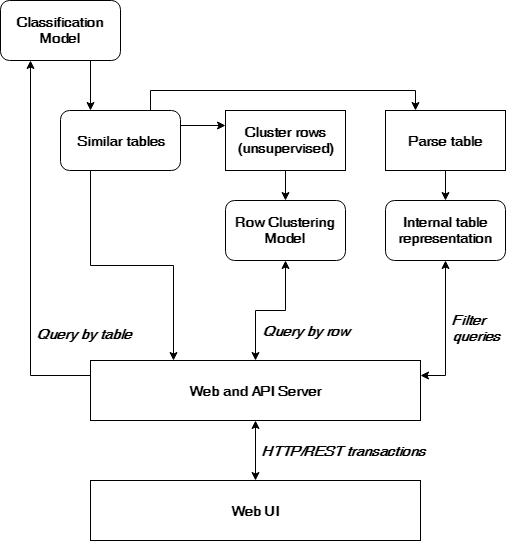}
\caption{Pipeline of operations facilitating interactive querying and analysis of classified tables}
\label{fig:query_pipeline}
\end{figure}

\subsubsection{Query by Table - Comparison of related tables}

To find similar tables from a given table, we perform clustering by following an approach of supervised nearest neighbour classification. 


The insights recorded from the neural network classifier in terms of probability vectors are used as features for calculating distance in a K-Nearest Neighbours classification model. This model acts as the core mechanic in the layer we develop for querying tables similar to a given table. The probability distribution from softmax layer of the deep-learning network is used directly as distance vectors. 
As demonstrated later in the paper, this model enables querying with (the vectorized representation) of a table to retrieve top similar tables, ideally belonging to the same company and standard type.

An interactive comparison and querying of tables is provided through a web interface served by a backend server which relies on a Nearest Neighbour classification model (See Figure \ref{fig:query_pipeline} on page \pageref{fig:query_pipeline}). 

\subsubsection{Query by Rows - Comparison of similar rows}

We can increase the utility of the system for retrieving similar tables by providing table-row level comparisons using the information we already have. Once the neighbours for a given table entry are obtained, the individual rows from all the tables are sliced, and converted to word embedding vectors. These word embedding vectors are then used to train an unsupervised Nearest Neighbour clustering model. This model enables us to query similar rows across the candidate tables. Mappings are stored to identify which row belongs to which table, so that this can be cross-referenced later for visualisation in the interface. 

For a selected row from the query table, the top-N similar rows are queried from across all the neighbouring tables, and highlighted in the interface (See Figure ~\ref{fig:row_grads_right} on page ~\pageref{fig:row_grads_right}). As a result, we are able to compare and analyse similar content within similar tables.

\subsubsection{Range Filters}
In addition to querying by row and retrieving similar rows, further filters are enabled on top of this result. The table structure itself is parsed into an internal representation which can be queried with custom row and column filtering queries. For instance, result rows can be filtered by numerical queries (See Figure \ref{fig:row_grads_filter_right} on page \pageref{fig:row_grads_filter_right}), and columns can be filtered by year (See Figure \ref{fig:row_grads_filter_year_right} on page \pageref{fig:row_grads_filter_year_right}). Although this querying works independently of row similarity, the two set of results can be juxtaposed in the final visualisation.

\section{System Implementation} 

We have built a browser-based application with a user interface and visualisation to interact with the classification and query results. In this section, we summarise the implementation technology and techniques, then introduce the main features of the system.

\subsection{Tools and Technologies}

\subsubsection{Data pre-processing}
The first step in the pipeline of processing PDF documents is to convert them into HTML markup using available tools. We were interested in obtaining an HTML representation where the PDF tables are be represented using \texttt{<table>...</table>} tags. Among the conversion tools tested, Adobe Acrobat\footnote{Acrobat DC, \url{https://acrobat.adobe.com/au/en/acrobat/acrobat-pro-cc.html}} was able to consistently preserve the semantics of table content and translate them into HTML table markup tags, while also preserving the correct structure in almost all the cases. The tool does not provide any batch conversion or command line access, hence the conversion part is manual.

\subsubsection{Classification}
The neural network architecture for table classification was built using TensorFlow from Google\footnote{TensorFlow, \url{https://www.tensorflow.org/tutorials/}}. The graph for classification is run once to produce the representation of sample tables as normalized probability distribution vectors.

\subsubsection{Table and Row Similarity}
The distance vectors obtained from classification step are used to train a KNN classifier implemented in Scikit-learn \cite{ScikitNeighbours}, which forms the basis to query similar tables from a given whole table.

Furthermore, individual rows from a set of similar tables are clustered using the unsupervised Nearest Neighbor implementation from Scikit-learn.

\subsubsection{Custom Querying}
We have a complete HTML table markup available from our initial conversion of the PDF document. While extracting tables for the dataset, the full markup is preserved. This information is utilised to provide additional functionality in the interface. The table markup is parsed to build a Python data structure representing the rows and columns of the table. Individual cells are parsed to identify numerical and text data types. Additionally, the first N rows of a table is parsed to look for possible column headers, most commonly the year corresponding to the numerical data. At the end of this processing, we have a backend data structure which contains rich information about the table, and we can run filtering queries on this.

\subsubsection{Server and Web Interface}
The application runs a Flask web server with Jinja templates\footnote{Armin Ronacher, Python Flask, \url{http://flask.pocoo.org/}}.  
Once a table or row based query is sent to the backend, the results are computed by the machine learning and query modules and result details are injected in the Jinja template as Javascript datastructures. The frontend scripts then perform the tasks of rendering this information and also providing the interactivity. REST interfaces are provided to the backend server to accept queries about numerical ranges and columns corresponding to specific years. 

\subsection{Overview of the Features}

\begin{figure*}
\includegraphics[width=7in]{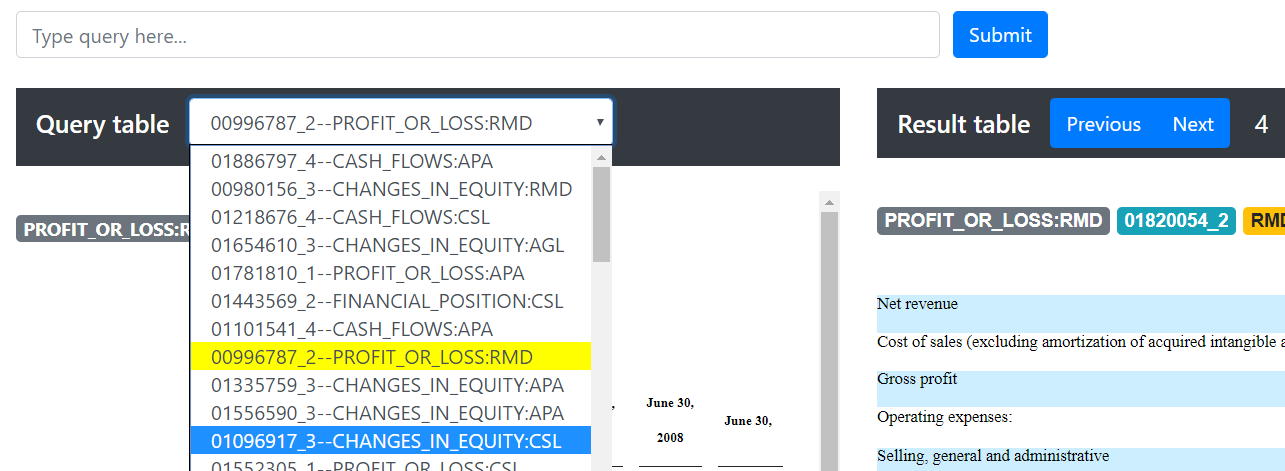}
\caption{Interface for selecting a table to query with. The image also shows the input to enter custom queries.}
\label{fig:table_sel}
\end{figure*}

\begin{figure*}
\includegraphics[width=7in]{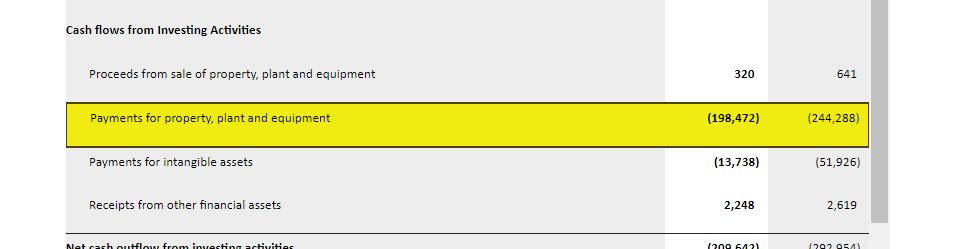}
\caption{Querying related rows in result tables. Clicking on any row in query table will highlight it and search for similar rows in similar tables.}
\label{fig:row_grads_left}
\end{figure*}

\begin{figure*}
\includegraphics[width=7in]{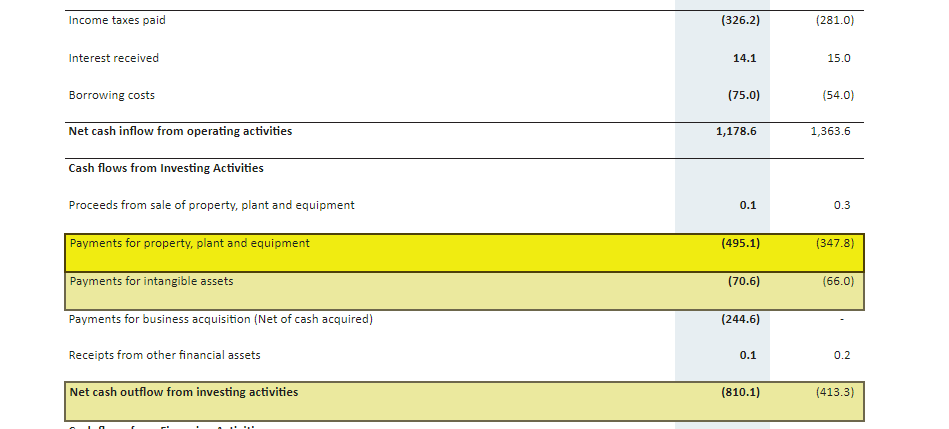}
\caption{Result of the query from \ref{fig:row_grads_left}. The most similar result is highlighted in bright yellow and other similar but distant results are highlighted in lighter yellow on the right side. Hovering on result rows also displays the actual distance from the query row from left side.}
\label{fig:row_grads_right}
\end{figure*}

\begin{figure*}
\includegraphics[width=7in]{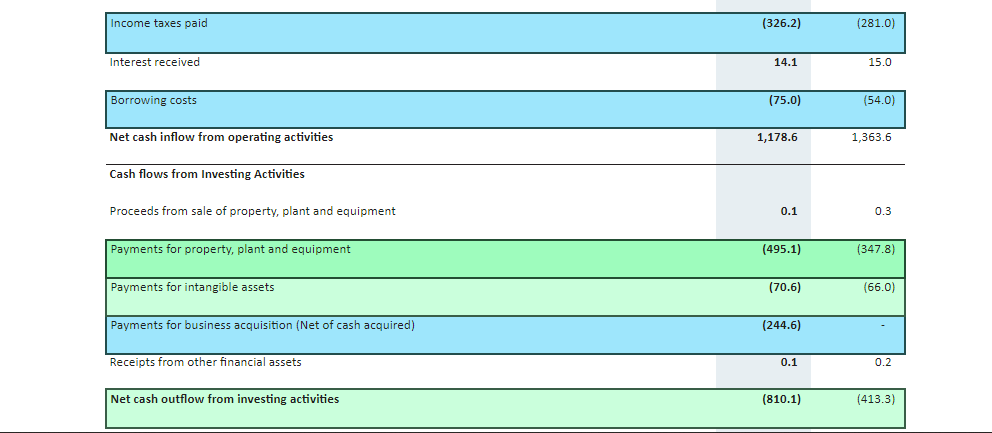}
\caption{Row filter results for the query \texttt{`gt 50 and lt 1500'} overlaid on row similarity results. The rows which contain any cell satisfying the numerical range query is highlighted in blue. This can work together with row query results. When the numerical query result intersects with row similarity results, the rows are highlighted in green, with contrast gradient to denote varying distance/relevance. Year column results are highlighted in purple.}
\label{fig:row_grads_filter_right}
\end{figure*}

\begin{figure*}
\includegraphics[width=7in]{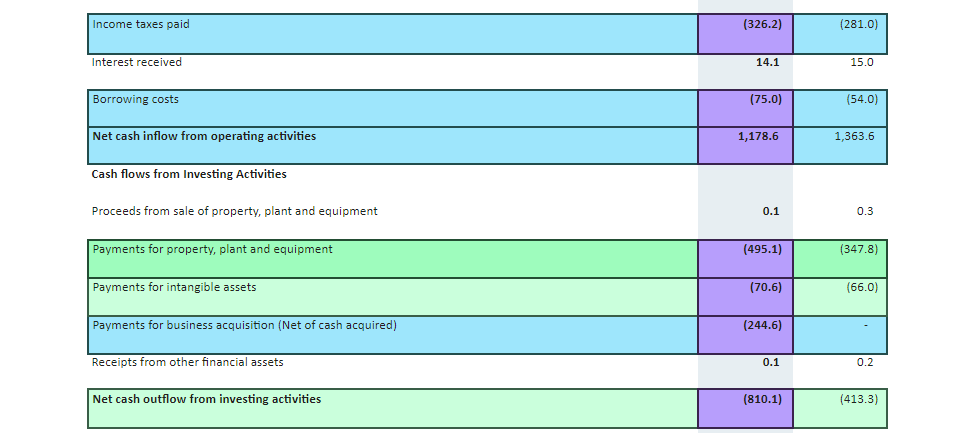}
\caption{Row filter results for the query \texttt{`gt 50 and lt 1500 and year 2016'}. Year column results are highlighted in purple.}
\label{fig:row_grads_filter_year_right}
\end{figure*}




The web interface provides the option to query by table with a drop-down selection of data set samples (See Figure \ref{fig:table_sel} on page \pageref{fig:table_sel}). Once the table is selected, top 5 most similar tables are queried from the backend model and displayed on the right side. User can browse through these tables using navigation controls.

The interface also reproduces the original visual style of the tables, thereby preserving any highlights and emphasise provided on headers or particular rows. This is enabled by cross-referencing the tables back to their original converted HTML files, extracting style information from those files, and combining them back together in the interface. The flexibility of Cascaded Style Sheets in applying styles in a modular fashion is the factor which enables this feature.

A row in the query table can further be analysed by clicking on it, which highlights similar rows in the result tables. The results are the most similar rows, and the actual distance can be viewed by hovering on the result rows. The highlighting emphasises the varying similarity of rows (See Figure \ref{fig:row_grads_left} on page \pageref{fig:row_grads_left} and Figure \ref{fig:row_grads_right} on page \pageref{fig:row_grads_right}).

This filtering based on queries can be juxtaposed with row similarity queries as seen in Figure ~\ref{fig:row_grads_filter_right} on page ~\pageref{fig:row_grads_filter_right}. The numerical range queries filter out rows with entries which falls in the range, and the columns corresponding to any query year is highlighted on top of this. The filter query can be an open range such as \texttt{gt 20}, or a closed range like \texttt{gt 20 and lt 500}. 

The year query if applied together with a range query enhances the results by highlighting the column in just the result rows (See Figure ~\ref{fig:row_grads_filter_year_right} on page ~\pageref{fig:row_grads_filter_year_right}).

\section{Evaluation}
We have evaluated the processing pipeline for its performances on  classification and query tasks.

\subsection{Performance of Classification}

\subsubsection{Dataset}
The annual report PDF documents were obtained from the publicly downloadable content in Australian Stock Exchange. A total of 146 tables were extracted from various reports, spanning across 5 different companies. For training the neural net, 80\% of the samples were used through random selection. Rest of the 20\% was used as testing set. In order to compare results between different word embeddings, the random state used was serialized and loaded in subsequent runs so that the same train-test split could be obtained.

\subsubsection{Methodology and Metrics}

During the training of the neural network classification model, the accuracy on test set prediction is evaluated using TensorBoard. The test accuracy is computed for a randomly sampled test batch on each iteration and the average overall accuracy is tracked. After the training, the entire test batch is run through the final model and the final predictions are stored. This is then used to compute the precision and recall values as follows:
\begin{displaymath}
    precision = \frac{TP}{TP + FP}
\end{displaymath}

\begin{displaymath}
    recall = \frac{TP}{TP + FN}
\end{displaymath}

where $TP$ is the number of true positives in the prediction, $FP$ the number of false positives, and $FN$ the number of false negatives. As we are dealing with multiple labels in the dataset, precision and recall is computed for each label and their average is computed weighted by the number of true positives for each label.

\subsubsection{Results}


\begin{figure*}
\includegraphics[width=0.9\linewidth]{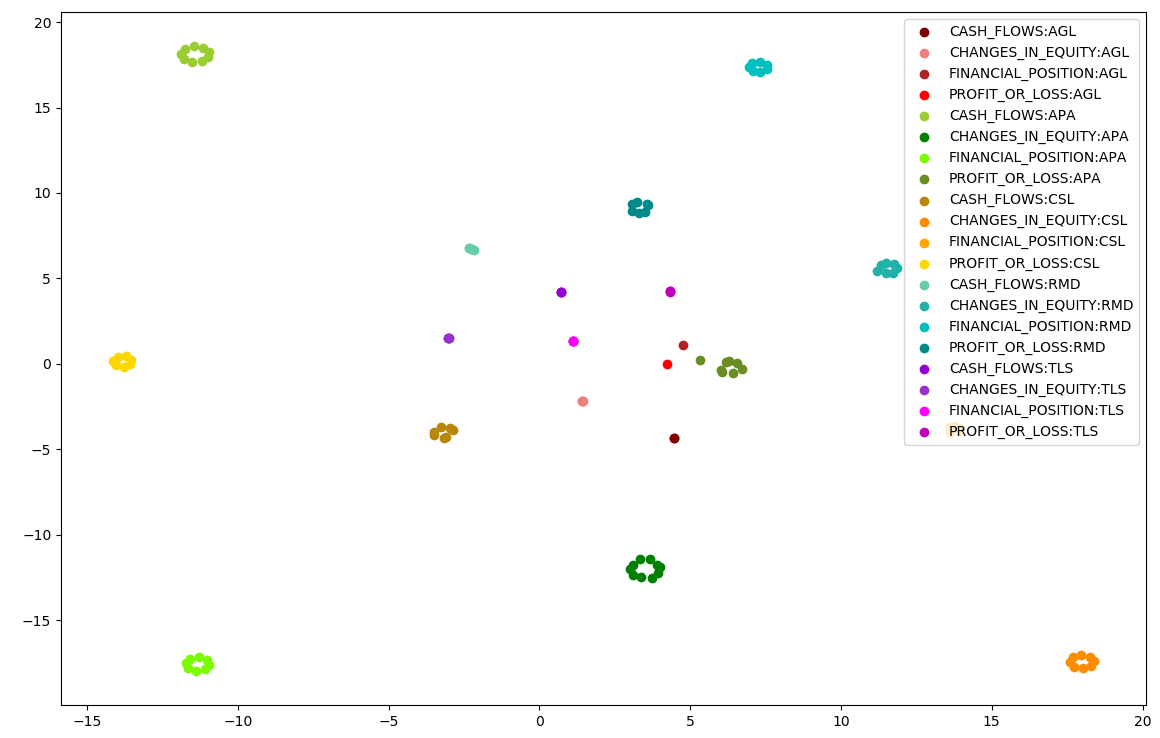}
\caption{Clustering of table types using custom word embedding}
\label{fig:tsne_custom}
\end{figure*}

\begin{figure*}
\includegraphics[width=0.9\linewidth]{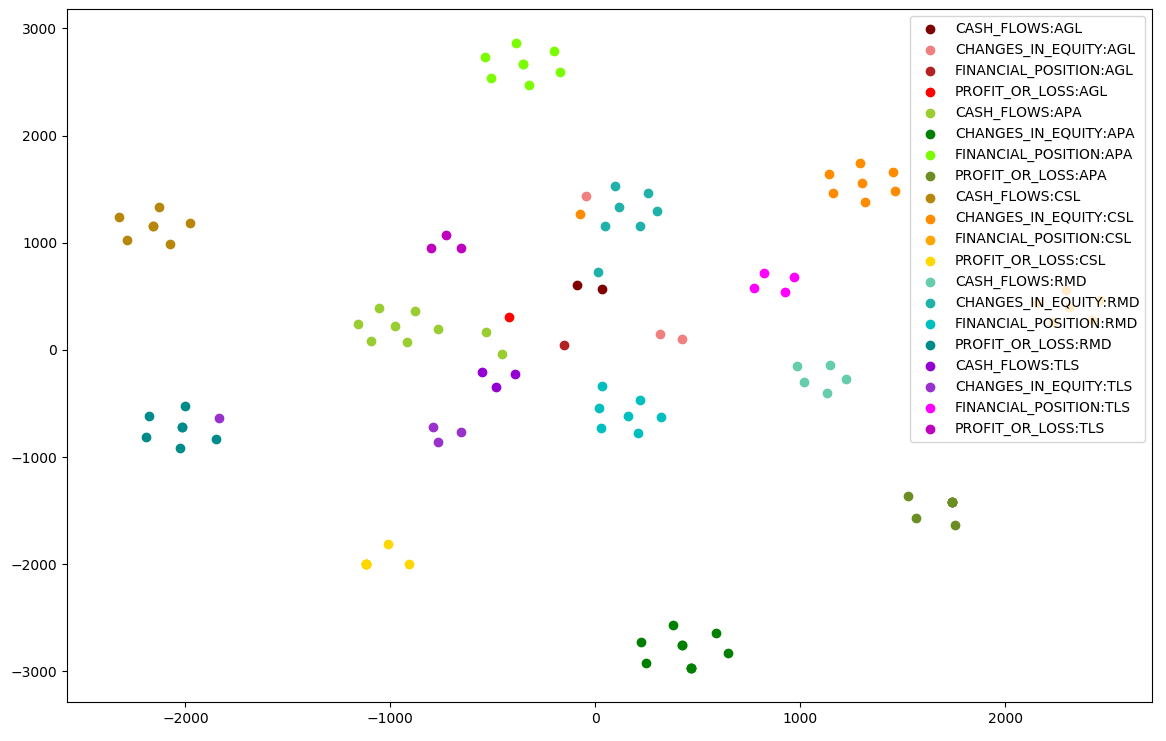}
\caption{Clustering of table types using Google News word embedding}
\label{fig:tsne_google}
\end{figure*}

The results obtained in classification using both the word embeddings - one from the custom word embedding trained on sample set vocabulary and another resulting from training based on Google News word2vec model - showed similar performance. Across various runs with slightly varying accuracy, both methods yielded an average of \texttt{95.45\%} overall accuracy, with precision and recall of \texttt{0.975} and \texttt{0.95} respectively.

We obtain two sets of distance vectors for the tables from the two different word embeddings used for training. Similar to their comparable performance in the neural network model, the clustering shows clear separation of different classes they are annotated with. This is visualised in the 2D plots generated by applying t-SNE algorithm on the distance vectors, for custom word embedding (See Figure ~\ref{fig:tsne_custom} on page ~\pageref{fig:tsne_custom}) and Google News embedding (See Figure ~\ref{fig:tsne_google} on page ~\pageref{fig:tsne_google}). The KNN model trained on custom word embedding showed \texttt{95.45\%} accuracy on average while the model based on Google News embedding showed \texttt{90.9\%} accuracy.

Results when trained on only table type (without any company information) showed that the clustering still retrieves similar tables from same company, but the top results also contain same table type from other companies. 

\subsection{Performance of Query}
\subsubsection{Methodology and Metrics}
The performance of querying by row (i.e., returning similar rows) is measured by manual inspection of the similarity results. To do this, we have established a ground-truth for each test query in terms of which other rows in the dataset should be considered similar to the query. 

For a given table, every row is iteratively chosen in the web interface, and the results of row similarity query is validated against the ground-truth. A count is kept of how many rows returned are ``valid'' results, and they are marked as hits. The misses are where the row query returned no reasonably similar result. The hit/miss rates are averaged over 
the test queries.

PostgreSQL is an open source relational database system. It includes a module called \texttt{pg\_trgm}, which enables string similarity search based on trigram matching \cite{PostgresSim}. For performance comparison, the tables are programatically parsed and row headers are inserted into PostgreSQL tables. The same pairs of tables compared in the web interface are paired up in the relational database, and for each row in the first table, similar rows are queried from the second table using string similarity search provided by the database. The generated results for each pair is verified manually and tallied in a similar way as we did in the web interface.

The hit rate $H$ for a given table is calculated as:

\begin{displaymath}
    H = \frac{N_h}{N_t} x 100
\end{displaymath}

where $N_h$ is the number of hits, and $N_t$ is the total number of rows iterated in the table.

\subsubsection{Results}

\begin{figure*}
\includegraphics[width=0.9\linewidth]{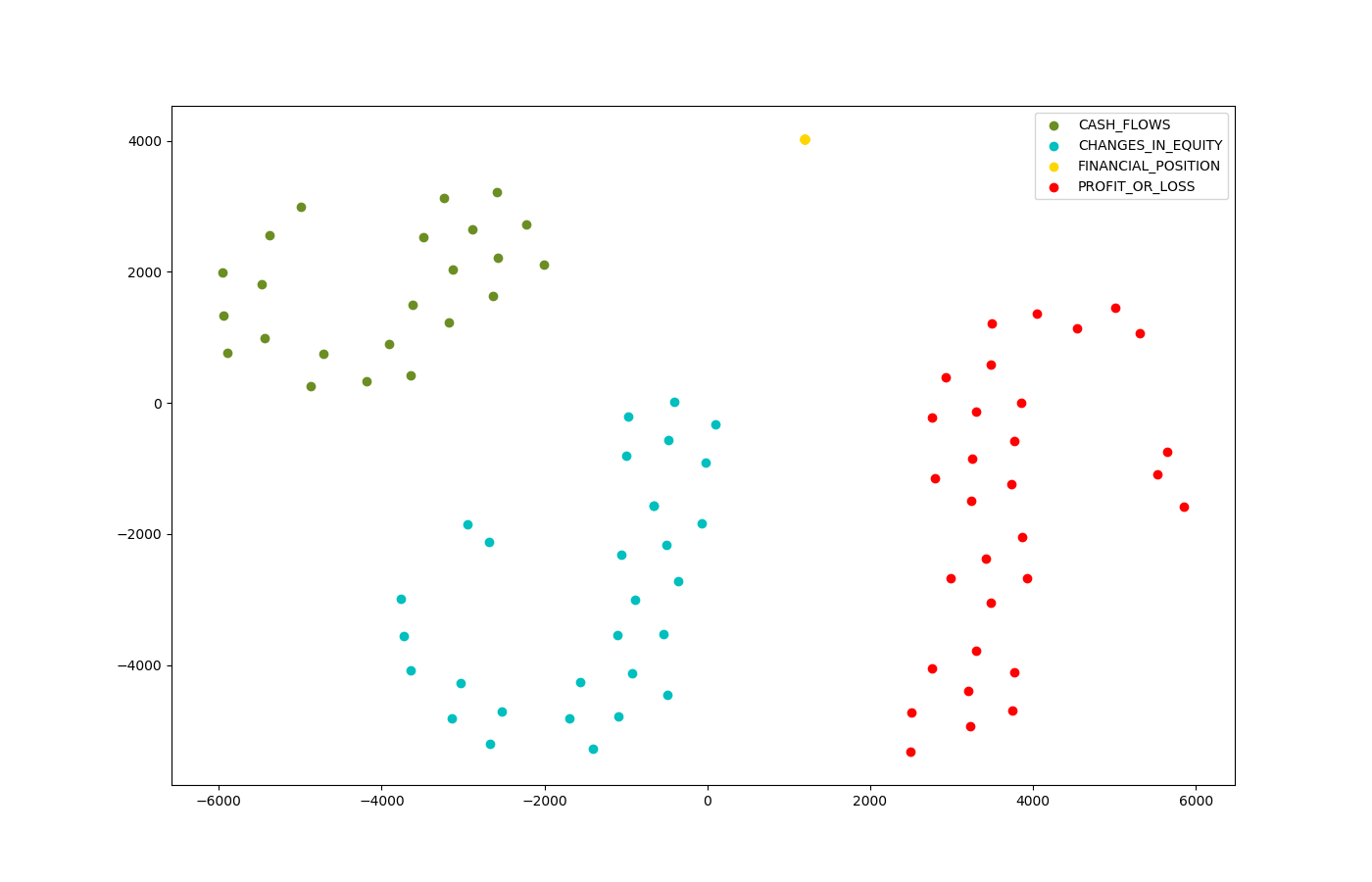}
\caption{Clustering of samples after training only on table type, without company information in label}
\label{fig:tsne_dummy}
\end{figure*}

\begin{table}
  \caption{Accuracy of querying similar rows with different search methods}
  \label{tab:rowsim}
  \begin{tabular}{ccl}
    \toprule
    Similarity metric&Average Hits\\
    \midrule
    Custom word embedding & 83.15\% \\
    Google News word embedding & 87.84\%  \\
    PostgreSQL string similarity & 75.08\%  \\
  \bottomrule
\end{tabular}
\end{table}

\begin{table}
  \caption{Hit rate of finding valid similar rows for a subset of table samples}
  \label{tab:rowsim_detail}
  \begin{tabular}{cccl}
    \toprule
    Sample ID & Google News & Custom embedding & PostgreSQL\\
    \midrule
    1 & 97.44\% & 87.18\% & 84.62\% \\
    2 & 91.67\% & 87.5\% & 83.33\%\\
    3 & 90\% & 85\% & 95\%\\
    4 & 76.19\% & 66.67\% & 90.48\%\\
    5 & 85.71\% & 85.71\% & 50\%\\

  \bottomrule
\end{tabular}
\end{table}

Although we observed similar performance on classification tasks using either of the word embeddings of choice, the row similarity measure is where Google News embedding outperforms the custom word embedding. This is possibly due to the fuzziness involved in searching for similar rows relying on a small set of word tokens. The comparative scores of both embeddings for row similarity checks on a set of tables can be found in Table ~\ref{tab:rowsim} on page ~\pageref{tab:rowsim}. Row similarity search based on word embedding vectors and clustering outperformed PostgreSQL's similarity search.

In the breakdown of individual sample in Table \ref{tab:rowsim_detail} on page \pageref{tab:rowsim_detail}, we can observe that for some tables, PostgreSQL displays a better performance than the word embeddings. This owes to the particular mix of vocabulary in the row headers. In the word embedding case, similarity of two rows is measured by first converting them into vectors of real numbers based on the embedding, and then computing the geometric distance. If most of the words contained in the two sentences are close to each other in the sample space, a sentence with seemingly different words may be picked up as a closer candidate. In case of PostgreSQL, similarity is strictly based on the actual words and letters which appears in the sentence. This explains the better performance of the relational query in some cases, although the machine learning method shows better overall performance.

For table type classification without company labels, our method has high accuracy. Figure \ref{fig:tsne_dummy} on page \pageref{fig:tsne_dummy} visualises the distribution of tables of each type in the model transformed space. There are clear margins between each type of tables represented by different colours.

\section{Conclusions}
In this paper, we presented a methodology and prototype implementation of a deep-learning based table processing pipeline architecture that provides quick browsing and querying access to similar table data across many PDF files.

The results and final implementation of this study shows how a practical system can be developed around a heterogeneous and unstructured data source like PDF documents. The degree of insights that can be obtained here without any hard-coded or static parsing of data can be appreciated more when we consider the manual effort which usually goes into retrieving information from varied PDF reports. 

The results are promising in providing a framework for using easily available public domain data for quick analysis, without any human effort or subscription to premium sources. The goal of achieving not precise search but similarity based queries presents a useful tool of analysis without the overhead of formal data ingestion processes.

The current implementation can be fitted to run with an API-based conversion tool to transform PDF files to HTML. This will allow us to incorporate the deep learning models with a complete pipeline starting from converting a bulk collection of PDF documents, processing them at once and making the results available through an interactive interface.
Combined with an automated system for retrieving and updating a large set of PDF documents as required, this pipeline could be of practical use when quick integration and simple inspection of the data is desired.




\bibliographystyle{ACM-Reference-Format}
\bibliography{project-bibliography}

\end{document}